\documentclass[aps,prc,twocolumn,superscriptaddress,showkeys]{revtex4-2}

\usepackage{amsmath,amssymb,bm}
\usepackage{graphicx}
\usepackage{multirow}
\usepackage{color}
\usepackage[normalem]{ulem}
\usepackage{hyperref}
\usepackage{CJKutf8}
\usepackage{subcaption}
\usepackage{ulem}

\newcommand{\eq}[1]{\begin{equation} #1 \end{equation}}

\makeatletter
\def\p@subsection{}
\makeatother

\begin{document}
	
	\begin{CJK*}{UTF8}{gbsn}
		
		
		\title{Role of tensor terms of the Skyrme energy-density functional on
			neutron deformed magic numbers in the rare-earth region}
		
		
		
		\author{Kai-Wen Kelvin-Lee (李 凯 文)}
		\affiliation{Department of Physics, Faculty of Science, Universiti Teknologi Malaysia, 81310 Johor Bahru, Johor, Malaysia}
		
		\author{Meng-Hock Koh (辜 明 福)}
		\email{kmhock@utm.my}
		\affiliation{Department of Physics, Faculty of Science, Universiti Teknologi Malaysia, 81310 Johor Bahru, Johor, Malaysia}
		\affiliation{UTM Centre for Industrial and Applied Mathematics, 81310 Johor Bahru, Johor, Malaysia}
		
		\author{L. Bonneau}
		\email{bonneau@cenbg.in2p3.fr}
		\affiliation{CENBG, UMR 5797, Université de Bordeaux, CNRS, F-33170, Gradignan,
			France}
		
		\date{\today}
		
		\begin{abstract}
			The role of the tensor part of the nuclear interaction is actively
			investigated in recent years due to experimental advancement yielding
			new data in nuclei far from the $\beta$-stability line. In this
			article we study the effect of this part of the nuclear
			interaction on deformed neutron magic numbers in the rare-earth region within 
			the Skyrme energy-density functional for various TIJ
			\cite{lesinski:2007} parametrizations. Two quantities signaling magic numbers 
			are considered: two-neutron separation energies and single-particle
			energies. They are calculated in isotopic series involving
			well-deformed rare-earth nuclei ranging from $Z=64$ to $Z=72$
			in the $N=100$ region. Obtained results show that, whereas the neutron-proton 
			tensor contribution to binding energies is important to reproduce neutron
			sub-shell closure at $N = 104$ in heavier rare earths Yb ($Z=70$) and
			Hf ($Z=72$) isotopes, like-particle tensor also plays a role in the
			single-particle spectrum around Fermi level and is even favored in
			lighter Gd ($Z=64$) and Dy ($Z=66$) rare-earth isotopes. 
		\end{abstract}
		
		\maketitle
		
		
	\end{CJK*}
	%
	%

	\section{Introduction}
	The discovery of gravitational-wave signal GW170817 coming from binary
	neutron stars (BNS) merger shows that BNS merger is more likely the
	candidate site for rapid neutron capture process (known as the
	\textit{r} process) \cite{Ligo:2017}. The nucleosynthesis through
	\textit{r} process occurs through rapid capture of free neutrons
	forming neutron-rich elements away from the beta-stability line. In
	the solar abundances, the \textit{r} process is responsible for the
	second and third peaks around $A = 130 \sim 138$ and $A = 195 \sim
	208$, respectively \cite{Arnould:2017}. These peaks are attributed to
	neutron shell closures giving rise to relatively high stability. A
	small peak is also observed around $A = 165$ in the rare-earth
	region. This peak was explained in terms of increasing nuclear
	deformation which stabilizes the nucleus similarly to the role of
	neutron closed shell \cite{surman:1997}. Substantial effort has been
	made by the experimental nuclear physics community to uncover this
	so-called deformed magic numbers. However, as we will see shortly,
	these magic numbers remain somewhat elusive.
	
	On the experimental side, considerable effort has been made resulting 
	in many proposals for new deformed magic numbers. In 1999, Asai and
	collaborators showed two minimum in the first $2^+$ energies in
	Dy isotopes \cite{asai:1999}. They proposed that the second
	minima at $N = 104$ coincides with the location of maximum deformation. 
	The first minimum at $N = 98$ was however dismissed as arising due to
	some local effect which enhanced the deformation around this
	isotope. This conclusion on $N = 98$ was also agreed upon in the work
	of \cite{soderstrom:2010} about a decade later based on systematic
	studies on yrast levels of Dy isotopes and the $4^+ \rightarrow
	2^+$ transition in $^{170}$Dy. Subsequent work by Patel
	\textit{et al.} \cite{patel:2014} in 2014 showed that a deformed magic
	number exists for neutrons at $N = 100$ in elements with proton number
	$Z \leq 66$, namely in Nd ($Z=60$), Sm ($Z=62$), Gd ($Z=64$) and
	Dy ($Z=66$) isotopes. This magic-number character of $N = 100$ in this
	region was, however, challenged three years later by Wu and
	collaborators \cite{wu:2017}. They reported to find no evidence of
	deformed subshell gap at $N = 100$ from the analyses of $\beta$- decay
	half-lives of Pm ($Z=61$) isotopes. Instead, they proposed two different
	magic numbers at $N = 96$ for $Z = 58$ to $Z=62$ and $N = 104$ for $Z
	= 63$ to $Z=66$. The more recent works of Hartley and 
	collaborators~\cite{hartley:2018,hartley:2020}
	on the other hand showed that $N = 98$ could instead be a candidate
	neutron subshell closure around $Z = 64$, contradicting the findings
	of Wu et al.~\cite{wu:2017} that $N = 104$ should be the deformed
	magic number in this isotope. Going to heavier rare-earth nuclei,
	\cite{watanabe:2019} reported emergence of new sub-shell closure at $N
	= 108$ in $_{72}$Hf ($Z=72$), W ($Z=74$) and Os ($Z=76$) isotopic series.
	
	On the theoretical side, there are rather limited studies to uncover
	the possible deformed magic numbers in the rare-earth region. To the
	best of our knowledge, all of the studies supported the magicity of $N
	= 100$ in the light rare-earth elements Sm and/or 
	Dy~\cite{satpathy:2004,ghorui:2012,kaur:2020}. The possibility of
	different subshell closures in heavier rare-earth isotopes ($Z>66$) as
	indicated by some experimental data was not explored. There were however
	calculations on $K$ isomers for example the $6^+$ in
	$^{170}\rm{Dy}$~\cite{Regan:2002,Rath:2003} and yrast levels in
	Dy isotopes~\cite{Yadav:2002}. Interestingly, Yadav et
	al.~\cite{Yadav:2002} reported that $N = 102$ is more likely the magic
	number in Dy isotopes instead of $N = 104$ based on energies of
	the ground-state and first $2^+$ states obtained within cranked
	Hartree-Fock-Bogoliubov calculations.
	
	One of the current major theme in nuclear theory is related to the
	impact of tensor two-nucleon interaction. While pioneering work on
	tensor effective potential was performed in 1977 by Stancu et
	al.~\cite{stancu:1977}, there was not much follow up of this work
	until mid 2000s when access to exotic nuclei was made possible through
	technological and experimental breakthroughs. Within the mean-field
	approach based on Skyrme energy-density functional, efforts have been
	made to design new parametrizations through either a perturbative or a full
	fitting procedure. In the perturbative approach, only the two zero-range
	tensor terms are adjusted while all other Skyrme parameters are
	kept constant. This is the case for the SIII+tensor
	(SIII+T) parameter sets of Refs.~\cite{stancu:1977,brink:2018} and 
	the SLy5+tensor (SLy5+T) parametrization of Ref.~\cite{colo:2007}. On
	the other hand, a fit of all parameters has been performed by Lesinski et
	al.~\cite{lesinski:2007} yielding a set of TIJ parametrizations which were
	applied to the investigations of spherical
	nuclei~\cite{lesinski:2007}, nuclear deformation~\cite{bender:2009} 
	and time-odd systems~\cite{Hellemans:2012}. Investigations on fit
	protocols of tensor effective potential components was also studied by Zalewski et
	al.~\cite{Zalewski:2008} who proposed that the single-particle levels
	should be considered instead of the usual bulk properties like the
	binding energies. 
	
	Within the Gogny mean-field approach, similar effort has been made by
	Anguiano et al.~\cite{Anguiano:2012} highlighting the need for
	inclusion of tensor effective potential. In this work a density-independant,
	finite-range tensor interaction term is added to the D1S
	parametrization, yielding the parametrization called D1ST2a. In this
	perturbative approach, independant like-nucleon and neutron-proton
	contributions are present in the effective two-nucleon potential, as
	in the zero-range Skyrme tensor potential, and the fitting
	protocol--involving neutron single-particle energies of $1\rm f_{5/2}$
	and $1\rm f_{7/2}$--yielded a strength of $-20$~MeV for the
	like-nucleon term and a much larger strength of 115~MeV for the
	neutron-proton term. Subsequently Grasso and
	Anguiano~\cite{Grasso:2013} studied the appropriate range for the strength of 
	the tensor terms within the Skyrme and Gogny energy density
	functionals (EDF) while Ref.~\cite{grasso:2014} showed that tensor effective potential
	is important to explain magicity at $N = 32$ and $N =34$ in the
	$^{52}$Ca and $^{54}$Ca nuclei. More recently, Bernard and
	collaborators~\cite{Bernard:2020} investigated the role in fission of the
	tensor terms of the Gogny EDF through a thorough comparison
	of several fission-related quantities--ranging from fission-barrier
	heights and paths to fission-fragment neutron emission--obtained with
	the D1S and D1ST2a parametrizations. One of the most important
	conclusion is that the added tensor terms are able to account for the
	new compact-symmetric fission configuration experimentally observed
	during the 2012 SOFIA campaign at GSI Darmstadt~\cite{Chatillon:2019}.
	
	From the rich literature showing that a tensor effective potential affects
	the single-particle levels ordering, 
	we are interested to investigate if inclusion of such a potential within our Skyrme EDF would allow us
	to explain deformed magic numbers suggested by experiment in the rare-earth region.
	We shift our attention to the heavier rare-earth nuclei  
	which have not gained much attention from theorists as compared to their lighter counterparts
	with particular interest in the $N = 104$ sub-shell closures.
	
	After a brief presentation of the relevant theoretical ingredients, we
	address successively in sections III to V the effect of the Skyrme
	tensor effective potential on charge quadrupole moments, two-neutron
	separation energies, and single-particle spectra. We give concluding remarks
	in section VI.

	\section{Theoretical approach}
	
	We considered several Skyrme fully refitted TIJ parametrizations namely
	\begin{itemize}
		\item T22, T24 and T26 (pure like-particle coupling)
		\item T22, T42 and T62 (pure neutron-proton coupling)
		\item T41 and T44 (mixed coupling).
	\end{itemize}
	The TIJ forces are labelled in such a way that $I$ and $J$ values are related to the proton-neutron $\beta$, 
	and like-particle $\alpha$ tensor coupling, respectively \cite{lesinski:2007}
	\begin{flalign}
		\alpha =& 60 (J - 2) \: \mbox{MeV fm}^3	\notag	\\
		\beta =& 60 (I - 2) \: \mbox{MeV fm}^3	\notag
	\end{flalign}
	with $\alpha = \alpha_C + \alpha_T$ and $\beta = \beta_C + \beta_T$.
	The subscript C and T refer to the central and tensor contributions, respectively.
	
	We have also included the original SIII \cite{beiner:1975} and SLy5 \cite{chabanat:1998} parametrizations
	and their counterparts in which tensor effective potential components are added perturbatively
	for comparison.
	The seniority force is used to approximate the residual pairing interaction
	whereby the neutron and proton pairing strengths were adjusted such that the BCS pairing gap yields
	the empirical Jensen formula \cite{jensen:1984}.
	The single-particle wave function is expanded on a deformed harmonic oscillator basis with
	a basis size of 16. The oscillator parameters $b$ and $q$ have been optimized
	to yield the lowest ground-state energy for each nucleus \cite{hafiza:2019}.
	We have limited ourselves to axial and parity symmetric nuclear shapes.
	
	Within the Skyrme EDF, in addition to the strength parameters of the central and
	spin parts of the effective potential, the tensor parameters $t_e$ and $t_o$ 
	enter the total binding energy in the $B_{14}$, $B_{15}$, $B_{16}$ and $B_{17}$
	coupling constants given by 
	\begin{flalign}
		&B_{14} = -\frac{t_1 x_1 + t_2 x_2}{8} + \frac{1}{4} (t_e + t_o)	\notag	\\
		&B_{15} = \frac{t_1 - t_2}{8} - \frac{1}{4}(t_e - t_o)	\notag	\\
		&B_{16} = -\frac{3}{8} (t_e + t_o)	\notag	\\
		&B_{17} = \frac{3}{8} (t_e - t_o).	\notag	
	\end{flalign}
	
	In order to isolate the contribution from the tensor parameters to the binding energy, 
	we separate the contributions coming from $B_{14}$ and $B_{15}$
	into two parts, such that:
	\begin{flalign}
		E_{B_{14}}^{C}& = -\Big(\frac{t_1 x_1 + t_2 x_2}{8} \Big) \mathbf{J}_{\mu \nu} \mathbf{J}_{\mu \nu}	\notag	\\
		E_{B_{14}}^{T}& = \frac{1}{4} (t_e + t_o) \sum_{\mu,\nu = x}^{z} \mathbf{J}_{\mu \nu} \mathbf{J}_{\mu \nu}	\notag	\\
		E_{B_{15}}^{C}& = \Big( \frac{t_1 - t_2}{8} \Big) \sum_{\mu,\nu = x}^{z} \mathbf{J}_{q,\mu \nu} \mathbf{J}_{q, \mu \nu}	\notag	\\
		E_{B_{15}}^{T}& = -  \frac{1}{4}(t_e - t_o) \sum_{\mu,\nu = x}^{z} \mathbf{J}_{q,\mu \nu} \mathbf{J}_{q, \mu \nu}	\notag
	\end{flalign}
	where $\mathbf{J}_{\mu \nu}$ is the spin-current density with $\mu, \nu = \{r,z,\phi\}$
	and $\mathbf{J}_{q,\mu \nu}$ is the spin-current density for the charge state $q$
	(see Ref.~\cite{Engel:1975} for their definition).
	The contributions from the $B_{16}$ and $B_{17}$ terms to the binding energy are
	\begin{flalign}
		E_{B_{16}} =& B_{16} \sum_{\mu=x}^{z} \Big(\mathbf{J}_{\mu \mu}\Big)^2 	\notag	\\
		E_{B_{17}} =& B_{17} \sum_{\mu=x}^{z} \Big(\mathbf{J}_{q, \mu \mu}\Big)^2 .	\notag 
	\end{flalign}
	
	Separating these terms in such a way allows us to draw out the contribution
	of the tensor part alone from all other non-tensor related terms to the binding energy.
	This means that the binding energy can be partitioned into
	\eq{E = E_{C} + E_{T}	\notag
	}
	where
	\eq{E_{C} = E_{kin} + E_{Coul} + E_{pair} +E_{B_x} + E_{B_{14}}^{C} + E_{B_{15}}^{C}
		\notag
	}
	with the contribution to the $E_{B_x}$ term comes from all the Skyrme coupling constants except
	for $B_{14}$, $B_{15}$, $B_{16}$ and $B_{17}$.
	The contribution to the $E_{T}$ comes from terms related to the tensor effective potential parameters $t_e$ and $t_o$ such that
	\eq{E_{T} = E_{B_{14}}^{T} + E_{B_{15}}^{T} + E_{B_{16}}^{C} + E_{B_{17}}^{C}.
		\notag
	}
	
	\begin{figure*}[h]
		\begin{centering}
			\includegraphics[width=1.0\textwidth]{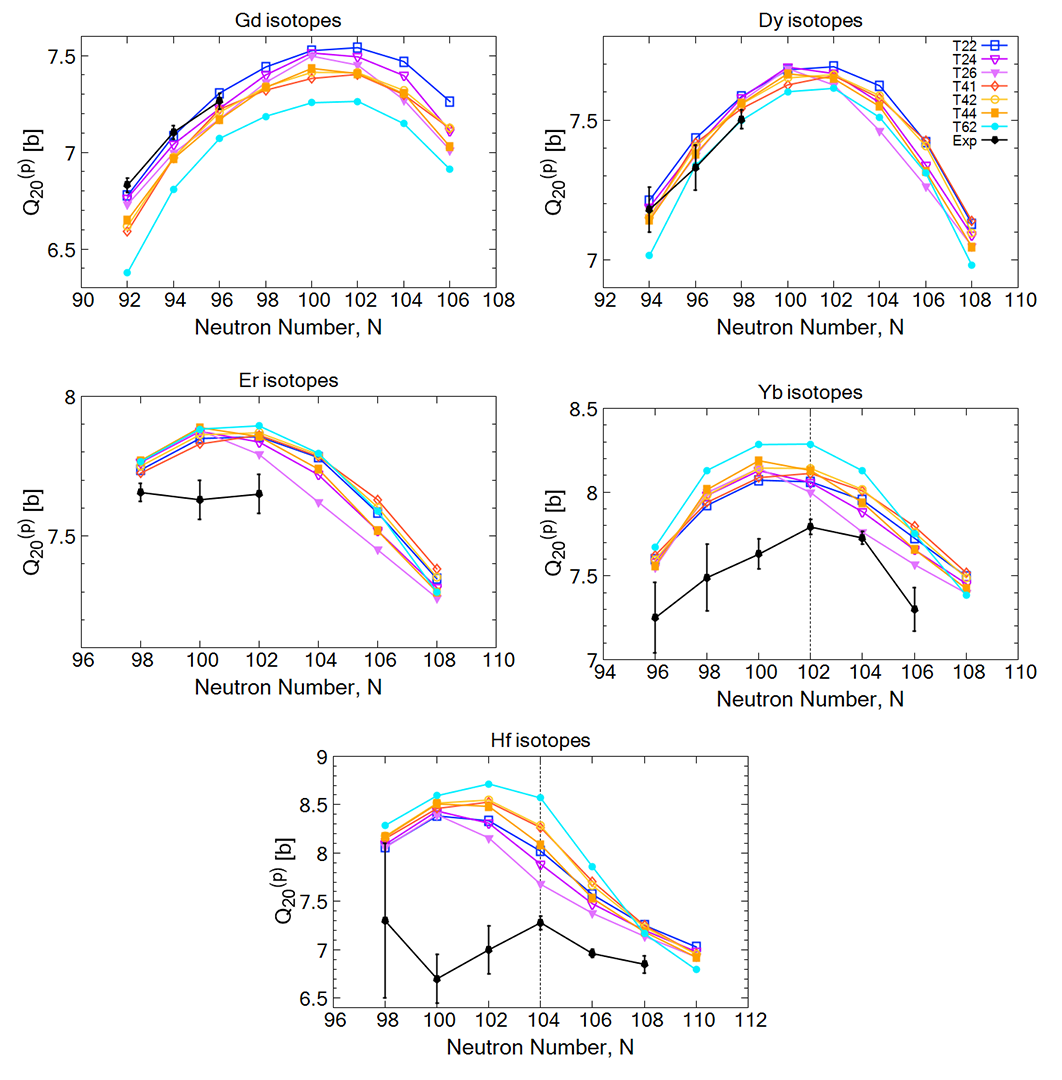}
			\par\end{centering}
		\caption{Intrinsic charge (proton) quadrupole moment obtained with seven TIJ parametrizations. 
			Experiment data are taken from \cite{Raman:2001}.}
		\label{fig Q20}
	\end{figure*}
	
	\section{Charge intrinsic quadrupole moment}
	
	We first present our ground-state intrinsic charge quadrupole moment for isotopic series 
	of $_{64}$Gd, $_{66}$Dy, $_{68}$Er, $_{70}$Yb and $_{72}$Hf and compared to experiment
	\cite{Raman:2001} in Figure~\ref{fig Q20}. Calculations with the various TIJ forces
	give good agreement with available experimental data. More importantly, we find a peak
	around $N \sim 100$, which corresponds to the maximum deformation in the region.
	The enhanced stability around this neutron number suggests the existence of deformed magic numbers, 
	giving us confidence that our investigation should be centered around $N\sim 100$.
	Yet, the charge quadrupole moment being a bulk property of the nucleus, it does not allow us
	to comment more on what are the possible predicted deformed magic numbers nor on the actual role of
	tensor effective potential. As such, we shift our attention to another observable expected to be
	more sensitive to a shell gap in the neutron single-particle spectrum, namely the two-neutron
	seperation energy.
	
	\section{Two-neutron separation energies}
	
\begin{figure*}
		\begin{centering}
			\includegraphics[width=1.0\textwidth]{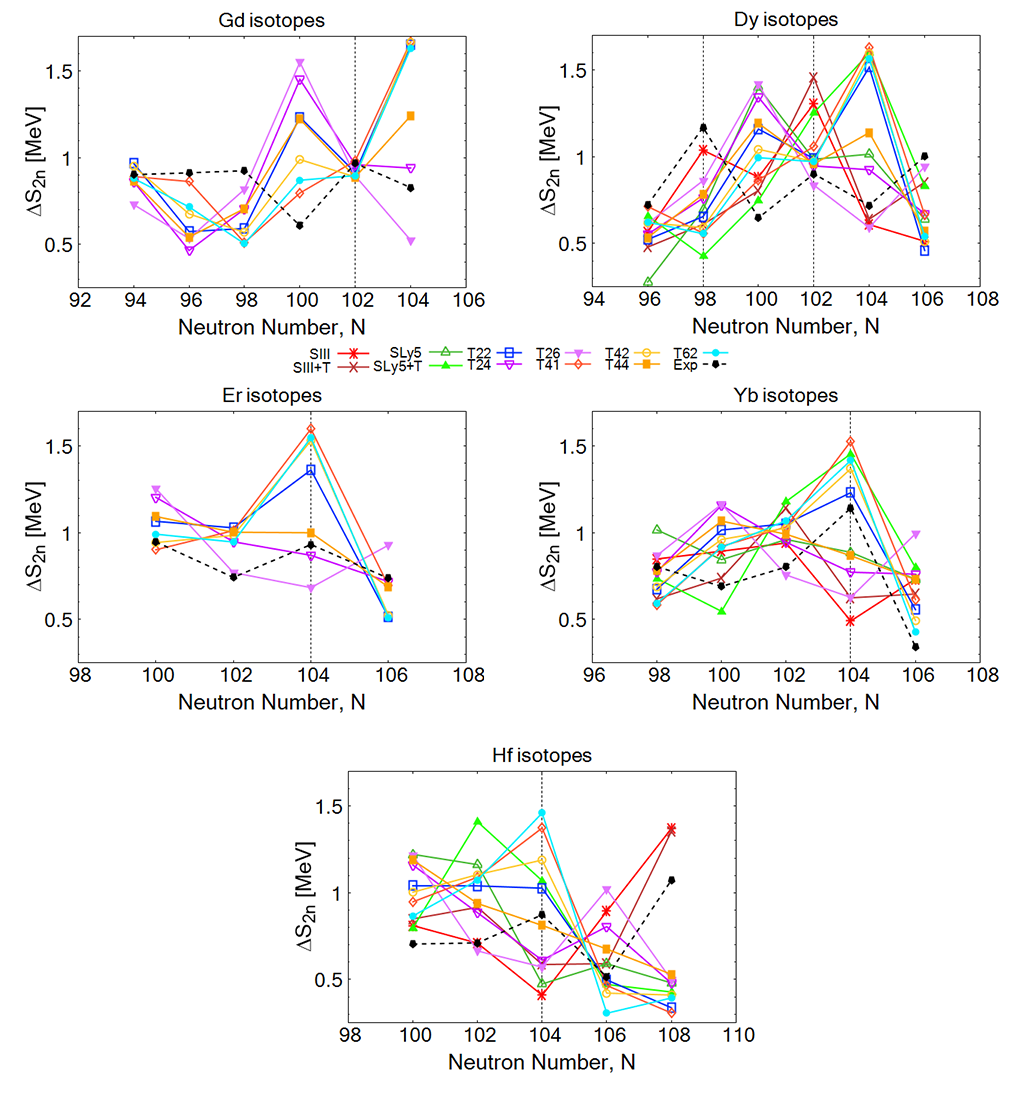}
			\par\end{centering}
		\caption{Two-neutron separation energy differential for 
			$_{64}$Gd, $_{66}$Dy, $_{68}$Er, $_{70}$Yb and $_{72}$Hf isotopes as a function of neutron number.}
		\label{S2n_diff}
	\end{figure*}

	We compute the two-neutron separation energy $S_{2n}$ and
	two-neutron separation energy differential $\Delta S_{2n}$ using the expression
	\eq{S_{2n} = E(N-2, Z) - E(N,Z)	\notag
	}
	\eq{\Delta S_{2n} = S_{2n} (N,Z) - S_{2n}(N+2,Z).	\notag
	}
	The calculated $\Delta S_{2n}$ are plotted in Figure~\ref{S2n_diff}
	together with experimental data taken from AME2016 \cite{wang:2017}.
	
	Let us first discuss the results for the three heavier elements considered in our
	study namely $_{68}$Er, $_{70}$Yb and $_{72}$Hf. The experimental data show a peak
	at $N = 104$ \cite{wang:2017} in these elements. To compare the theoretical results
	with data, we take the T22 as the reference parametrization because it is such that
	$\alpha = \beta = 0$, although the Skyrme parameters $t_e$, $t_o$ are not zero. 
	The T22 parameter set manages to produce a pronounced peak at $N = 104$ especially in $_{68}$Er
	and $_{70}$Yb. The peak at this neutron number is even more enhanced when increasing
	$\beta$ by considering the T42 and T62 parametrizations. This shows that neutron-proton
	tensor coupling constant $\beta$ is essential to reproduce the neutron $N = 104$
	sub-shell closure in these rare-earth nuclei. This behavior of the $N=104$ peak
	with $\beta$ is even more marked in $_{72}$Hf isotopes.
	
	On the contrary, increasing like-particle tensor coupling constant $\alpha$
	with a vanishing $\beta$ contribution, i.e. in the sequence T22 $\rightarrow$
	T24 $\rightarrow$ T26, results in larger dips, instead of peaks, at $N=104$.
	With non-vanishing $\beta$ and still increasing $\alpha$ in the sequence
	T41 $\rightarrow$ T42 $\rightarrow$ T44 parametrizations, we see that the
	pronounced peak at $N = 104$ obtained with T41 decreases when using the T42
	parametrization, and then vanished totally with T44. This clearly shows that
	like-particle tensor coupling tends to remove the $N = 104$ peak in heavy
	rare-earth nuclei. Therefore the reproduction of this peak requires small $\alpha$
	values and positive, sizeable $\beta$ values.
	
	Concerning the two parametrizations obtained from perturbative fits of
	the tensor effective potential, we find that the SLy5+T improves the results
	as compared to the original SLy5 parametrization. Indeed a significant peak is
	found with SLy5+T in the $\Delta S_{2n}$ plot at $N = 104$ for $_{70}$Yb isotopes
	instead of a minute peak at $N = 102$ with SLy5. However, neither SLy5 nor SLy5+T
	are able to reproduce the magicity of $N = 104$ in $_{72}$Hf isotopes.
	
	Before moving on to lighter elements of the rare-earth region, we make a remark on the
	$\Delta S_{2n}$ at $N = 108$ in $_{72}$Hf isotopes. In this element, the experimental
	point at $N = 108$ is higher than the one at $N = 104$. We did not manage to reproduce
	this pattern in our calculations. In spite of this, we see that the $\Delta S_{2n}$
	between $N = 106$ and $N = 108$ exhibits a positive slope when using T42 and T62 forces,
	while all other TIJ forces give a negative slope. This reinforces the conjecture
	that neutron-proton tensor coupling is more favored in heavy rare-earth nuclei
	and can, at the very least, reproduce the experimental trend qualitatively.
	
	In $_{66}$Dy isotopes, peaks are seen at $N = 98, 102$ and possibly $N = 106$
	in experimental $\Delta S_{2n}$ \cite{wang:2017}. Calculations with TIJ parametrizations
	are not able to reproduce this experimental trend. Instead, the SIII and SIII+T
	parametrizations performed better there. The SIII parametrization generates peaks at
	$N = 98$ and $102$ while the SIII+T calculations yield a peak at $N = 102$ and
	follow the experimental trend at $N = 106$. Results with the TIJ parameter sets,
	however, yield two peaks at $N = 100$ and $N = 104$. The former peak is enhanced
	when increasing the like-particle coupling constant $\alpha$, while the latter
	is more pronounced when the neutron-proton coupling constant $\beta$ is larger.
	Indeed, T22 calculations serving as a reference, we see that the peak
	at $N = 100$ is more pronounced when going to T24 and T26 forces.
	Conversely, the peak at the same neutron number is reduced when going from
	T22 to T42 and then to T62 forces. The reverse is seen at $N = 104$ where going
	from T22 to T24 induced a sharp drop in the $\Delta S_{2n}$.
	Comparing the results obtained with T41, T42 and T44 also indicates that 
	strong like-particle tensor coupling is undesirable to produce a peak at $N = 104$,
	similar to what is found in heavier rare-earth elements. Before closing this discussion
	of $_{66}$Dy results, we would like to draw the reader's attention to the fact
	that while our TIJ calculations do not reproduce experimental data of
	Wang et al.~\cite{wang:2017}, the TI2 results are however in agreement with Wu
	et al.~\cite{wu:2017} who showed that $N = 104$ forms a sub-shell closure.
	Clearly more experimental data in light rare-earth nuclei are needed to resolve this discrepancy.
	
	Finally, we comment on the results for the lightest rare-earth element considered in our study. 
	In the $_{64}$Gd isotopes, the experimental $\Delta S_{2n}$ is almost constant over the range
	94--98 of $N$. It makes a dip at $N = 100$ before forming a peak at $N = 102$.
	Our TIJ calculations fails to reproduce this trend, and the same pattern seen in the
	variation of the above $N = 100$ and $N = 104$ peaks with $\alpha$ and $\beta$ is obtained here.
	
	To conclude this section, we can say that the TIJ parametrizations produce two persistent
	peaks at $N = 100$ and $N = 104$. The peaks can be obtained in particular with the T22
	parameter set for which the central and tensor contributions cancel, yielding $\alpha = \beta = 0$.
	When switching to TIJ forces with $\alpha > 0$, we find an enhancement for the $N = 100$ peak
	while TIJ forces with $\beta > 0$ accentuates the peak at $N = 104$.
	
\begin{figure*}
		\begin{centering}
			\includegraphics[width=1.0\textwidth]{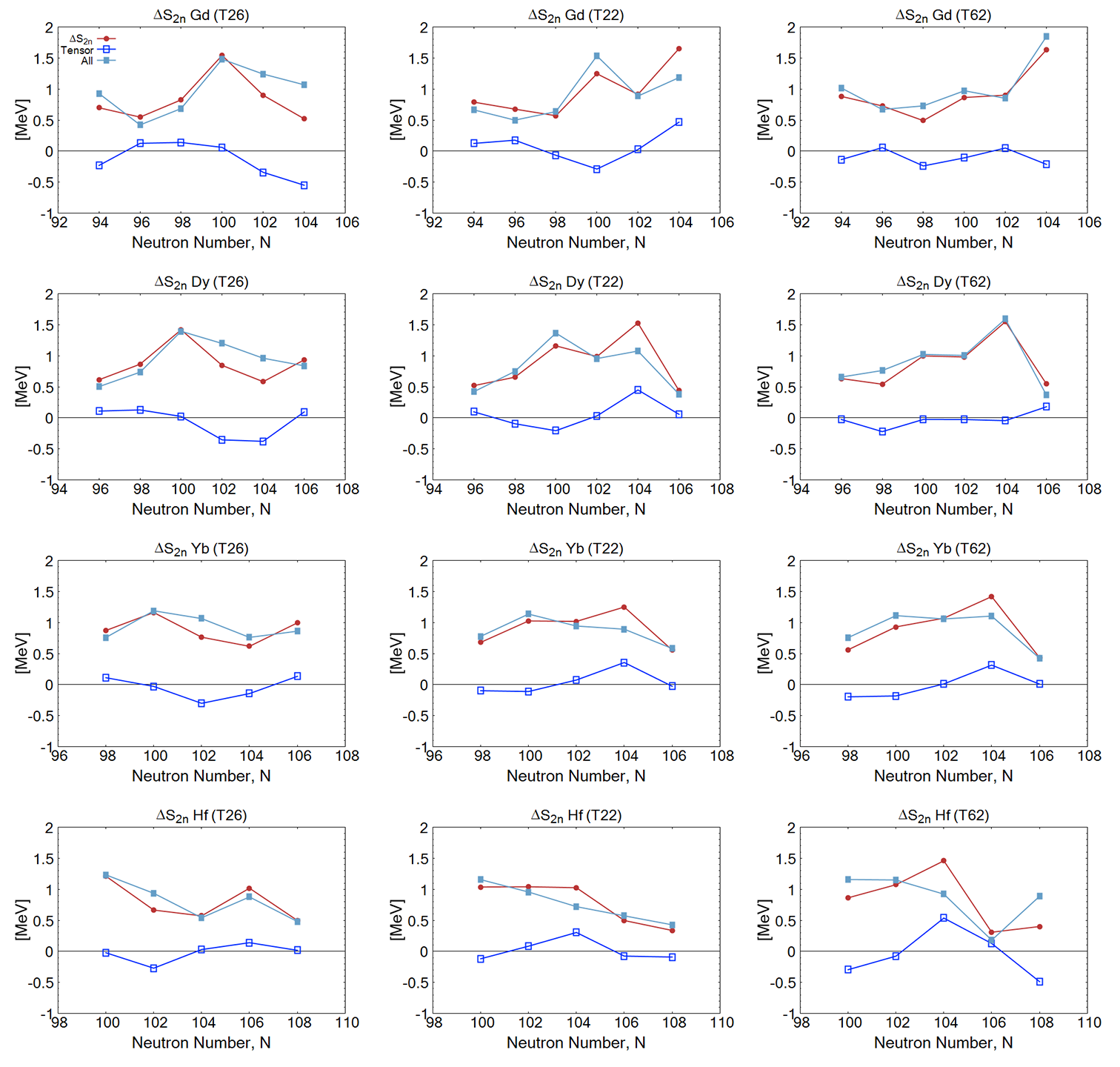}
			\par\end{centering}
		\caption{Contributions of non-tensor and tensor components to the $\Delta S_{2n}$.}
		\label{tensor vs all contribution}
	\end{figure*}

	In order to understand the role of the tensor effective potential, we plot in
	Figure~\ref{tensor vs all contribution} the contributions of $E_C$ and $E_T$ terms
	of the Skyrme energy-density to $\Delta S_{2n}$ as a function of $N$ for
	T26, T22 and T62 parametrizations. In all considered nuclei, the tensor contribution
	is small as compared to the sum of all other terms, but it plays an important role in
	shaping the fine structure of the patterns seen in Figure~\ref{S2n_diff}.
	
	In $_{72}$Hf, the tensor contribution is particularly crucial when using T22 and T62
	parameter sets. Indeed the contribution from all other terms to $\Delta S_{2n}$ does not
	yield a peak at $N = 104$, which can only be obtained thanks to the tensor component.
	
	In $_{70}$Yb, non-tensor terms alone do produce two peaks at $N = 100$
	and $N = 104$ with the T22 and T26 forces. When including tensor, the calculated
	$\Delta S_{2n}$ at $N = 100$ is decreased while the point at $N = 104$ is push
	upwards yielding only one peak at $N = 104$. A similar effect is observed in 
	in $_{64}$Gd and $_{66}$Dy elements.
	
	\section{Neutron single-particle energy spectra}
	
\begin{figure*}
		\centering
		\includegraphics[width=\textwidth]{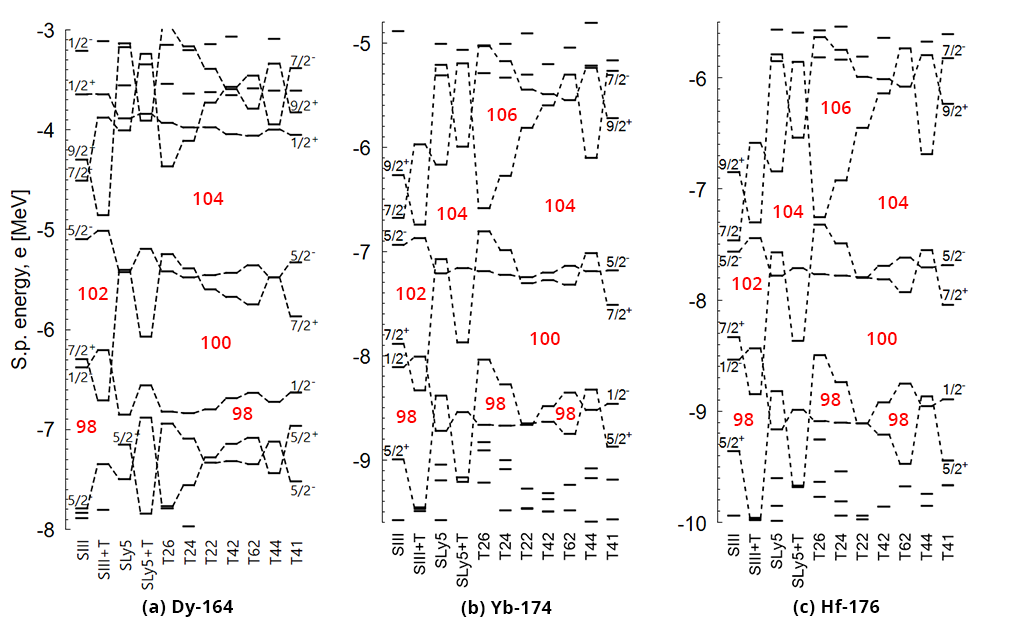}
		\caption{Neutron single-particle (s.p) spectra for (a)Dy-164, (b)Yb-174 and
			(c)Hf-176 as a function of Skyrme parameteriazations.}
		\label{sp_level}
	\end{figure*}

	We now turn our attention to the neutron single-particle levels for some nuclei
	in Figure~\ref{sp_level}. The variation in the calculated $\Delta S_{2n}$ with
	different Skyrme parametrizations in Figure~\ref{S2n_diff} coincide with the
	variation in the single-particle energy gap.
	
	In $_{66}$Dy isotopes, two pieces of information can be learnt.
	On the one hand pure like-particle ($\alpha$) tensor coupling (as in T24 and T26)
	favors sub-shell closure at $N = 100$. In fact, a substantial
	single-particle energy gap appears at $N = 96$ only with T26 force. Pure
	neutron-proton ($\beta$) tensor coupling, on the other hand, favors sub-shell closure
	at $N = 104$. The $N = 98$ sub-shell closure, while not reproduced by any TIJ forces,
	seems to be accounted for by a strong neutron-proton rather than like-particle
	tensor component. This is correlated with the decreasing trend in the single-particle
	energy gap at $N = 98$ when going from T24 to T26. However, according to Hartley et al in
	Ref.~\cite{hartley:2018}, ``$1/2$ neutron orbital above the $7/2$
	one is required to explain the decay properties of $^{162}_{\;\;63}$Eu''.
	This suggests to explore refinements to existing parametrizations.
	
	Let us move to the single-particle states of $_{70}$Yb and $_{72}$Hf isotopes.
	The peak in $\Delta S_{2n}$ at $N = 104$ is related to the widening of the
	single-particle energy gap between the $7/2^-$ and $7/2^+$ states seen in
	Figure~\ref{sp_level}. The energy gap between the two states increases with $\alpha$
	(along the sequence of calculations T22 $\rightarrow$ T42 $\rightarrow$ T62), while it 
	decreasing from T24 to T26, that is to say when $\alpha=0$ and $\beta$ increases.
	With T22, we see that the $7/2^+$ state is located below a $5/2^-$ state
	and remains almost at the same energy when switching on neutron-proton tensor coupling
	and keeping $\alpha=0$. In contrast the $7/2^-$ state keeps beeing shifted higher
	in energy, giving rise to a very large energy gap when increasing neutron-proton tensor
	strength.
	
	Moreover an important observation is made regarding the like-particle tensor coupling
	by comparing the results obtained with T42 and T44 forces. A slight increase of
	like-particle tensor coupling in T44 causes tremendous lowering of the $7/2^-$ state,
	while elevating the $7/2^+$ state above $5/2^-$. Consequently, the $N = 104$ is not a
	sub-shell closure for T44 while it is so for T42. This suggests that strong neutron-proton
	coupling is important to reproduce this deformed magic number in the heavy rare-earth region.
	However, the increasing energy gap at $N = 106$ sub-shell in $_{70}$Yb and $_{72}$ Hf isotopes
	with T24 and T26 could indicate the importance of like-particle tensor coupling.

	\section{Conclusion}
	
	In conclusion, we have performed Skyrme Hartree--Fock-BCS calculations for even-even
	rare-earth nuclei with $Z = 64$ up to $Z = 72$. We have found a maximum deformation
	around $N \sim 100$ which confirms that the neutron deformed magic numbers
	could be found in this neutron-number region.
	
	Then we have calculated two-neutron separation
	energies and studied their difference $\Delta S_{2n}$ between two consecutive even-$N$ values.
	with several TIJ parametrizations of the Skyrme energy-density functional.
	Two persistent peaks have been found at $N = 100$ and $N = 104$.
	These peaks have been obtained in all considered nuclei with the T22 parameter set (for which
	$\alpha=\beta=0$) except for $_{72}$Hf. While the $N = 100$ peak is even more pronounced
	when switching on like-particle tensor terms (driven by the $\alpha$ coupling constant)
	the $N = 104$ is enhanced by neutron-proton tensor terms (driven by the $\beta$ coupling
	constant). Comparison with experimental data of Ref.~\cite{wang:2017} suggests that
	neutron-proton tensor terms are favored in heavy rare-earth nuclei to reproduce
	the $N = 104$ peak. In contrast, increasing like-particle tensor strength with a fixed,
	positive $\beta$ coupling constant (in T41, T42, T44 parametrizations) has the
	detrimental effect of decreasing the $N=104$ peak. In the lighter rare-earth elements
	$_{64}$Gd and $_{66}$Dy however, the situation is not so clear. In these nuclei the like-particle
	tensor terms can produce a peak in $\Delta S_{2n}$ at some neutron numbers depending
	on the parametrization.
	
	To better understand the role of the tensor terms on this observable,
	we have studied the contribution to $\Delta S_{2n}$ arising solely from the $t_e$ and $t_o$
	parameters of the tensor effective potential and shown that, while being small, this
	contribution is important to produce the $\Delta S_{2n}$ peaks. We have also studied the
	neutron single-particle spectra for various parametrizations of the Skyrme EDF. A neat
	correlation between the peak structure of $\Delta S_{2n}$ and large single-particle energy
	gaps around Femi level could thus be evidenced.
	
	Overall the present work indicates that $N=104$ can be considered as a ``deformed'' magic
	neutron number thanks to neutron-proton tensor coupling in heavy rare-earth elements and that 
	like-particle tensor coupling is not desirable in this region. However, in lighter rare-earth elements
	the situation is less clear and further work is called for to better understand the intricate
	role of tensor terms of the effective nucleon-nucleon potential.

	\begin{acknowledgments}
		K.W K.L and M.H.K would like to acknowledge Malaysian Ministry of Education
		for the financial support through the Fundamental Research Grant Scheme \\
		(FRGS/1/2018/ST/G02/UTM/02/6)
		and UTM \\ (R.J130000.7854.5F028).
	\end{acknowledgments}

	\bibliography{mybibfile}
	

\end{document}